\documentclass[10pt]{article}
\usepackage[english]{babel}
\usepackage[figuresright]{rotating}
\usepackage{array}
\usepackage{graphicx}
\usepackage{pdflscape}
\usepackage{colortbl}
\usepackage{xcolor}
\RequirePackage{amsthm,amsmath,amsfonts,amssymb}

\usepackage{amsmath,amsfonts}
\usepackage{amsbsy}
\usepackage{color,soul}
\usepackage{verbatim}
\usepackage{setspace}
\usepackage{rotating}
\usepackage{array}
\usepackage{graphicx}
\usepackage{pdflscape}
\usepackage{booktabs}
\usepackage{colortbl}
\usepackage{xcolor}
\usepackage{multirow}
\definecolor{lightgray}{gray}{0.9}
\usepackage{xcolor}

\usepackage{enumerate}

\def\bc{\begin{center}}
\def\ec{\end{center}}

\title{Impact of COVID-19  on Air Quality in Israel}
\author{Sarit Agami\\
Department of Economics\\
Hebrew University, Mount Scopus, Jerusalem, Israel\\
email:sarit.agami@mail.huji.ac.il}

\begin{document}
\maketitle


\begin{abstract}
The COVID-19 pandemic has caused, in general, a sharp reduction in traffic and industrial activities. This in turn leaded to a reduction in air pollution around the world. It is important to quantity the amount of that reduction in order to estimate the influence weight of traffic and industrial activities over the total variation of air quality. The aim of this paper is to evaluate the impact of the COVID-19 outbreak on air pollution in Israel, which is considered one of the countries with a higher air pollution than other Western countries.
The results reveal two main findings: 1. During the COVID-19 outbreak, relative to its earlier closest period, the pollution from transport, based on Nitrogen oxides, had reduced by 40$\%$ on average, whereas the pollution from industrial, based on Grand-level ozone, had increased by 34$\%$ on average. Relative to 2019, the COVID-19 outbreak caused a reduction in air pollution from transport and industrial as well. 2. The explanation percent of the time period of COVID-19 is at most 22$\%$ over the total variation of each pollutant amount.
\end{abstract}



\section{Introduction}
Air pollution causes morbidity, death, and economic damage.
The sources of air pollution in Israel include man made air pollution sources such as transportation, power plants, factories, as well as natural sources, such as dust storms. In addition to the pollution that comes from Israel, the migration of pollution brings additional pollution from Europe or the desert. Other sources of pollution are domestic air pollution, soil pollution, quarries, and the result of pesticide spraying. Severe air pollution exists in the Bay of Haifa, Tel Aviv and Gush Dan as well as in Jerusalem. Haifa and its surroundings have air pollution from factories and transport alike, when the  air pollution is a product of a number of factories operating in a small area, producing high and unusual air pollution in terms of the quantities and types of toxins emitted into the air.  Common pollutants in Haifa Bay are Sulfur dioxide (SO$_{2}$), Nitrogen oxides (NO$_{x}$), Particulate matter (PM), and Volatile Organic Compounds (VOCs), such as Benzene.
In Tel Aviv and Gush Dan, air pollution is mainly from transportation; the area suffers from severe air pollution caused mainly by vehicles on which the transport is based in and around Tel Aviv-Yafo, as well as buses and trucks stuck in traffic jams, and airport. Transportation emits NO$_{2}$, NO$_{x}$, Grand-level ozone (O$_{3}$), Carbon monoxide (CO), and particulate matter 10 micrometers or less in diameter (PM$_{10}$).
\\
\indent The first tackle with the COVID-19 pandemic was reduction of the economic activity. Particularly a sharp reduction in road traffic, air traffic, shipping and industrial activities. This in turn leaded to a reduction in greenhouse gas emissions and air pollution around the world.
\noindent For example, one of the largest drops in pollution levels could be seen over the city of Wuhan, in central China, which was put under a strict lockdown in late January. The city of 11 million people serves as a major transportation hub and is home to hundreds of factories supplying car parts and other hardware to global supply chains. According to Nasa, NO$_{2}$ levels across eastern and central China have been 10\%-30\% lower than normal.
NO$_{2}$ levels also dropped in South Korea, which has long struggled with high emissions from its large fleet of coal-fired power plants but also from nearby industrial facilities in China.
\\
\indent In this paper we give an initial investigation of the influence of COVID-19 time period on air quality in Israel.
The global COVID-19  epidemic in Israel began to spread towards the end of February 2020. As part of the Israeli Ministry of Health's deal with the epidemic, starting in mid-March, regulations imposed on the public. At  first, a local closure on places where the virus is spreading had taken, as well as movement the public sector to the emergency and the private sector to a limited 70\% service.
Branches that worked in order - up to $100\%$, while reducing as much as possible, were, for example, the energy sector (including electricity, natural gas, oil, water), food industries (including agriculture, supermarkets, transport, storage and more), all freight and storage services, the ports and shipping companies, and workplace dealing with construction or infrastructure work. Industries that only some employers could work with $100\%$ were, for examples, various economic sectors, and economy and industry.
The public transport activity was also reduced to a quarter of its normal size, train traffic was disabled and the number of passengers in the taxi was limited.
It was decided to ban crowds of over 10 people, and to shut down all recreation and leisure.
Next, the regulations became tougher: people were ordered the avoidance of leaving home, except for emergency situations that required it. Later on, they asked leaving the place of residence no longer than 100 meters and for a short time.  On April 19, for the first time, easing restrictions applied to the public took effect.  In the beginning of May, relief was added (for example, people were allowed to go more than 100 meters).
Based on these dates, the restrictions time period is March 1 , 2020 until May 1, 2020. During writing this paper, the available data we had was until May 2, 2020. Therefore, throughout this paper, we refer to the time period of March 1 , 2020 until May 2, 2020 as the “COVID-19 period”. In order to examine any influence of this time period on air quality, needs to compare this period with its earlier closest period, with its parallel period a year ago, and with its parallel period over various years ago. If any influence exists, we should quantity its amount, and its relative weight over the total variation of a given pollutant, as well. In this paper, we consider the earlier closest period to be the period of January 1, 2020 until February 29, 2020. For the comparison with a year ago, we compare the whole period of January 1, 2020 until May 2, 2020, with the same period at 2019. Throughout this paper, we refer to the period of January 1 until May 2, for a given year, as the “full period”. We compare the full period over the years 2000-2020, as well.
\\
\indent This paper is organized as follows. Section 2 describes the data framework used to evaluate the air quality during the full period over the considered years. Section 3 presents the impact of the COVID-19 period on air quality, while Section 4 examines this impact along with weather variables. Section 5 gives a discussion and a summary of the analysis results.
\section{Data}
The data we use in this paper was downloaded from the Ministry of the Environment web \cite{ime20}. We consternate on the two regions of Haifa and Gush-Dan, which, as was mentioned in Section 1, have highly air pollution in Israel.
The list of the considered stations in Haifa and Gush-Dan, that includes their names, names in the paper, type, and area type, is described in Table 1. The general stations are stations located in a representative area, at the height of the roofs of buildings or in open rural areas. These stations are not near specific emission sources, such as industrial plants or roads.
The transport stations are highway stations along major transport routes. Measuring at this height makes monitoring of the transport stations to best represent the concentration of pollutants exposed to pedestrians, cafes and drivers in the city.
\begin{landscape}
\begin{table}[h!]
\caption{\textbf{List of Stations}}.
\centering
\fontsize{6.3}{0.96}\selectfont
\begin{tabular}{c|cccccccccccccccccc}
\\
&&&& \textbf{Haifa}  \\
\midrule
\textbf{Name - Full}&Kiryat Haim Regavim&Kiryat Shprintzak&Hogim&Ahuza General&Haifa Igud&Kfar Hasidim&Nave Shaanan\\
\midrule
\textbf{Name - in Paper}&Regavim&Shprintzak&Hogim&Ahuza&Igud&Hasidim&Shaanan \\
\midrule
\textbf{Station Type}& General&General&General& General&Industrial&General&General\\
\midrule
\textbf{Area Type}&Urban&Urban& Urban&Urban&Industrial&Rural&Urban\\

\midrule
\\
\midrule
\textbf{Name - Full}&Nesher&Park ha-Carmel&Kiryat Ata&Kiryat Binyamin&Kiryat Tivon&Kiryat Yam&Atzmaut
\\
\midrule
\textbf{Name - in Paper}&Nesher&Carmel&Ata&Binyamin&Tivon&Yam&Atzmaut\\
\midrule
\textbf{Station Type}& General& General&General&General&General&General&Traffic\\
\midrule
\textbf{Area Type}&Urban&Rural&Urban&Suburban&Suburban&Urban&Urban\\
\midrule
\\

\\
&&&& \textbf{Gush-Dan}  \\
\midrule
\textbf{Name - Full}&Remez&Rail Station Wolfson&Rail Station Komemiyut&Rail Station Yoseftal&Rishon Lezion&Amiel&Ironi Dalet&Levinsky
\\
\midrule
\textbf{Name - in Paper}&Remez&Wolfson&Komemiyut&Yoseftal&Rishon&Amiel&IroniD&Levinsky
\\
\midrule
\textbf{Station Type}&Traffic&Traffic&Traffic&Traffic&Traffic&Traffic&Traffic&Traffic \\
\midrule
\textbf{Area Type}&Urban&Indoor&Indoor&Indoor&Urban&Urban&Urban&Urban\\
\midrule
\\
\midrule
\textbf{Name - Full}&Kvish 4&Yefet Yaffo&Yad Avner&Holon&Hamashtela&Petah Tikva Road&Antokolsky&Ehad ha-Am \\
\midrule
\textbf{Name - in Paper}&Kvish4&Yefet&Avner&Holon&Mashtela&PT&Antok&Am \\
\midrule
\textbf{Station Type}&Traffic&Traffic&General&General&General&General&General&Traffic\\
\midrule
\textbf{Area Type}&Urban&Urban&Urban&Urban&Suburban&Urban&Urban&Urban\\

\bottomrule
\end{tabular}
\end{table}
\end{landscape}

The resolution of the data is half-hour in a day, and the considered air pollutants are PM (PM$_{10}$ and PM$_{2.5}$), NO$_{x}$, NO$_{2}$, NO, CO, O$_{3}$, SO$_{2}$, Benzene, Toluene, and Ethyl Benzene (EthylB). In addition, we use the weather variables wind-direction (WD), wind-speed (WS), temperature (Tmp), and relative-humidity (RH). We consider at each region the pollutants and weather variables having no more than 10\% missing values each one, at a specific station. In order to deal with unreasonable negative values of the pollutants, we converted negative values smaller than -1 to zero, while we treated as missing data the negative values that are larger than -1. The measurement units of each variable are described in Tables 2, 3.

\begin{table}[h!]
\caption{\textbf{Measurement Units of the Pollutants Variables}}.
\centering
\begin{tabular}{c|ccccccccccccc}
\midrule

Variable& PM$_{10}$&	PM$_{2.5}$&CO&	NO$_{x}$&NO$_{2}$& NO&O$_{3}$& SO$_{2}$ &Benzene&	Toluene&EthylB 	 \\
\midrule
Units	 &$\mu$g/$m^3$& $\mu$g/$m^3$&	ppm&	ppb&	ppb&	ppb&ppb&ppb&ppb&ppb&ppb\\
\bottomrule
\end{tabular}
\end{table}
\begin{table}[h!]
\caption{\textbf{Measurement Units of the Considered Weather Variables}}.
\centering
\begin{tabular}{c|cccc}
\midrule

Variable& WD&	WS&	Tmp&	RH	 \\
\midrule
Units	 &deg&	m/sec&	c degrees&$\%$\\
\bottomrule
\end{tabular}
\end{table}
\section{Impact of COVID-19}
We follow three comparisons in order to examine the influence of COVID-19 time period on air quality:
\\
1. A comparison of COVID-19 period (March 1, 2020 until May 2, 2020) with its earlier closest period.
 \\
2. A comparison of the full period (January 1 until May 2), over the years 2020 and 2019.
\\
3. A comparison of the full period (January 1 until May 2), over the years 2000-2020.
\\
Each comparison is based on means comparison of a given pollutant at a specific station. The calculation of the mean omitted the missing values. For testing the significant of the difference, we used the t-test for two means comparison.
In order to evaluate the relative contribution of the COVID-19 period to the total variation of a given pollutant, at a specific station, we used a linear regression model. This model includes an indicator variable, \emph{ind}, which is defined to be $0$ if the date of the record belongs to the earlier closest period, and $1$ if the date of the record belongs to the COVID-19 period. That is, this variable distinguishes between the COVID-19 period and the other period. Then, the model is of the form $Y=\alpha+\beta ind$, where $Y$ is the amount of a given pollutant at a specific station. The resulted $R^2$ of the model obtains the desired result. By this we can understand the explanation percent of the total variation of a specific pollutant as contributed by transport and industrial activates.

\subsection{Comparison of COVID-19 period with its earlier closest period}
Here we compare, for a given pollutant at a specific station, its mean at the COVID-19 period with its mean at the earlier closest period. Most of the pollutants had a reduction in their means at the COVID-19 period relative to the earlier closest period. But some of them had increasing. In most cases, the reduction/increasing was significant. We report on the significant results only. The relative change of each pollutant, which is defined as $(mean(covid-19)-mean(earlier))/mean(earlier))*100\%$, over the considered stations for Haifa and Gush-Dan, is summarized in Tables 4, 5. These tables contain the $R^2$ that resulted by the above regression model.
\\
\begin{landscape}
\fontsize{7.4}{1.16}\selectfont
\begin{table}[h]
\caption{\textbf{COVID-19 period (March 1, 2020 - May 2, 2020) vs. its earlier closest period (January 1, 2020 - February 29, 2020), Haifa}}.
\centering
\begin{tabular}{c|c|c|c|c|c|c|c|c|c|c|c|c|c|c|c}
\midrule

&Regavim&Shprintzak&Hogim&Ahuza&Igud&Hasidim&Shaanan&Nesher&Carmel&Ata&Binyamin&Tivon&Yam\\
\midrule
NO$_{2}$&-39.4&-35.64&-31.83&-36.62&-30.47&-30.54&-32.38&-25.69&-33.54&-21.8&-9.21&-33.65&-44.94 \\
\midrule
$R^2$&0.05&0.04&0.04&0.06&0.04&0.03&0.04&0.03&0.03&0.02&0.00&0.06&0.09 \\
\midrule
\midrule
No&-57.98&-51.00&-23.69&-59.14&-40.6&19.05&-28.91&-44.67&-23.28&-28.57&-3.91&-53.45&-71.9 \\
\midrule
$R^2$&0.01&0.01&0.01&0.04&0.01&0.00&0.01&0.02&0.00&0.00&0.00&0.04&0.04 \\
\midrule
\midrule
NO$_{x}$& -42.94&-38.53&-30.77&-41.19&-32.3&-27.07&-31.89&-30.13&-32.14&-22.48&-8.88&-37.17&-51.34\\
\midrule
$R^2$&0.04&0.04&0.03&0.07&0.03&0.02&0.04&0.03&0.03&0.02&0.00&0.06&0.08\\
\midrule
\midrule
SO$_{2}$& -53.47&&-3.24&-41.57&-27.69&9.08&-36.27&-23.8&-16.14&-24.31&25.21\\
\midrule
$R^2$&0.04&&0.00&0.01&0.01&0.00&0.02&0.01&0.03&0.00&0.00\\
\midrule
\midrule
Benzene&-49.5&&&&-44.9&&&&&&-39.13\\
\midrule
$R^2$&0.13&&&&0.19&&&&&&0.06\\
\midrule
\midrule
Toluene&-25.65\\
\midrule
$R^2$&0.01\\
\midrule
\midrule
EthylB&-55.83\\
\midrule
$R^2$&0.04\\
\midrule
\midrule
\rowcolor{lightgray} O$_{3}$&&13.77& 18.43&&27.69&22.6&14.89&14.71&15.27&24.78&&&35.56&&\\
\midrule
$R^2$&&0.09&0.18&&0.1&0.12&0.11&0.07&0.17&0.14&&&0.13\\
\midrule
\midrule
\rowcolor{lightgray}CO&&&&&&&4.96&&&-11.42&&&&&\\
\midrule
$R^2$&&&&&&&0.01&&&0.05\\
\midrule
\midrule
\rowcolor{lightgray}PM$_{2.5}$&&&&31.71&17.15&&18.53&15.08&32.62&12.56&-4.24&&30.75&&\\
\midrule
$R^2$&&&&0.02&0.01&&0.01&0.01&0.01&0.00&0.00&&0.02\\
\midrule
\midrule
\rowcolor{lightgray}PM$_{10}$&33.84&&&67.7&48.45&&62.36&60.95&&69.55&47.82&51.57&&&\\
\midrule
$R^2$&0.01&&&0.02&0.01&&0.01&0.01&&0.02&0.01&0.01\\
\midrule

\bottomrule
\end{tabular}
\end{table}
\end{landscape}

\begin{landscape}
\begin{table}[h]
\caption{\textbf{COVID-19 period (March 1, 2020 - May 2, 2020) vs. its earlier closest period (January 1, 2020 - February 29, 2020), Gush-Dan}}.
\centering
\fontsize{7.4}{1.16}\selectfont
\begin{tabular}{c|c|c|c|c|c|c|c|c|c|c|c|c|c|c|c|c|c|c|c|c|c}
\midrule
&Remez&Wolfson&Komemiyut&Yoseftal&Rishon&Amiel&IroniD&Levinsky&Kvish4&Yefet&Avner&Holon&Mashtela&PT&Antok&Am\\
\midrule
NO$_{2}$&-47.6&-17.31&-44.26&-44.98&-40.91&-30.94&-39.71&-49.02&-35.58&-43.99&-47.15&-36.4&-45.39&-48.19&-40.98&-37.7 \\
\midrule
$R^2$&0.17&0.01&0.09&0.06&0.11&0.08&0.09&0.22&0.10&0.09&0.10&0.05&0.12&0.14&0.08&0.12\\
\midrule
\midrule
No&-62.2&&-60.53&-54.53&-57.43&-36.46&-49.91&-66.04&-54.84&-62.65&-40.4&-30.66&-49.33&-63.83&-38.79&-51.53 \\
\midrule
$R^2$&0.04&&0.08&0.05&0.08&0.03&0.02&0.10&0.04&0.04&0.02&0.01&0.02&0.05&0.01&0.04 \\
\midrule
\midrule
NO$_{x}$& -54.84&&-57.05&-52.73&-48.57&-33.02&-43.6&-58.31&-44.46&-52.07&-45.15&-34.64&-46.16&-54.14&-40.47&-44.79\\
\midrule
$R^2$&0.08&&0.08&0.06&0.10&0.06&0.06&0.14&0.07&0.06&0.07&0.03&0.10&0.10&0.05&0.07\\
\midrule
\midrule
CO&-26.88&&&&-31.25&&&-50.31&-18.87&&-57.33\\
\midrule
$R^2$&0.07&&&&0.23&&&0.20&0.04&&0.14\\
\midrule
\midrule
SO$_{2}$&&&&&&&&&&&-8.6&&5.99&&-1.85 \\
\midrule
$R^2$&&&&&&&&&&&0.03&&0.01&&0.00\\
\midrule
\midrule
\rowcolor{lightgray}O$_{3}$&&&&&&&&&&40.64&36.75&53.36&54.21&&&&&&\\
\midrule
$R^2$&&&&&&&&&&0.13&0.13&0.17&0.22\\
\midrule

\midrule
\rowcolor{lightgray}PM$_{2.5}$&10.61&&&-16.67&26.81&&&&31.62&20.17&22.86&24.67&&&&&&&\\
\midrule
$R^2$&0.01&&&0.02&0.03&&&&0.03&0.03&0.02&0.03\\
\midrule
\midrule
\rowcolor{lightgray}PM$_{10}$&&&&&&&&&&&40.1&&&&&&&&\\
\midrule
$R^2$&&&&&&&&&&&0.03\\
\midrule
\bottomrule
\end{tabular}
\end{table}
\end{landscape}

A number of common trends are seen, as follows.

\subsubsection{Nitrogen oxide}
 The pollutants No, NO$_{x}$, and NO$_{2}$ are all compounds of Nitrogen oxide.
 Generally, the most common source of Nitrogen oxide air pollution is internal combustion engines for motor vehicles, and power plants. Particulary in Israel, the main source of nitrogen oxides is air pollution from land vehicles such as cars, buses and trucks. During the COVID-19 period, these pollutants of NO, NO$_{x}$, and NO$_{2}$ had a consistent reduction relative to the earlier closest period, over all the considered stations in Haifa and Gush-Dan. The reduction is, on average, -33.35 in Haifa, and -46.33 in Gush-Dan. Note that in Gush-Dan the reduction is higher than in Haifa; This is coincide with the fact that the main source of air pollution in Gush-Dan is from transport.
 In more details, the reduction for each component of these pollutants in Haifa is: NO$_{2}$: -31.21, NO: -36,
 NO$_{x}$ :-32.83, and in Gush-Dan: NO$_{2}$: -40.63, NO: -51.94,NO$_{x}$: -47.33.
The explanation percent of the total variation of each of these pollutants by the period indicator is ranged between 0 to 0.09 in Haifa, and between 0.01 to 0.22 in Gush-Dan. Particulary, the percent explanation of Nitrogen oxide by the period in Gush-Dan is a little higher than in Haifa, which again coincide with the above fact.

\subsubsection{Ground-level ozone (O$_{3}$)}
Ground-level ozone is a secondary pollutant created by photochemical reaction between primary pollutants such as nitrogen oxides, and hydrocarbons (volatile organic compounds) in the presence of solar radiation. These pollutants are emitted from industrial pollution and pollution from transport.
In contrary to Nitrogen oxide, O$_{3}$ had increased during the COVID-19 period relative to the closest period, over all the considered stations in Haifa and Gush-Dan. The increasing is, on average, 20.86 in Haifa, and 46.24 in Gush-Dan. This surprising result is due to increasing in promotion of transport and energy infrastructures \cite{globes}, and due to increasing in home renovations \cite{biznes}. That is, industrial pollution. The explanation percent of O$_{3}$ by the period indicator is ranged between 0.07 to 0.18 in Haifa, and between 0.13 to 0.22 in Gush-Dan.
\subsubsection{CO}
Carbon monoxide emissions usually occur as a result of incomplete combustion processes of natural gas, oil and coal. The main industrial activities contributing to carbon monoxide emissions are: metal processing, power generation, metal and coal mining, food production, gas and oil production, chemical production, cement and lime production, plaster and concrete production, oil production and refining.
Other sources of carbon monoxide emissions include human sources - transportation (vehicles, aircraft, ships), construction equipment, home heating, cigarettes and fires. And natural resources - volcanoes, forest burning, lightning.
One of the most important sources of massive exposure to carbon monoxide in humans is tobacco smoking.
During the COVID-19 period, CO had decreased relative to the earlier closest period over all the considered stations in Gush-Dan, by -36.93 on average. This can be due to the reduction in transport. The explanation percent of CO by the period indicator is ranged between 0.13 to 0.22.
But, in the two stations in Haifa for which CO was measured, one station had a reduction of -11.42 in CO, with $R^2=0.05$, whereas the other station had an increasing of 4.96, with $R^2=0.01$, both general stations. The last increasing can be due to increasing in industrial pollution, as was noted above.

\subsubsection{PM$_{10}$ and PM$_{2.5}$}
Particulate matter has natural and human sources. Natural sources that emit particles with volcanoes, dust storms, forest and grass fires, vegetative pollen, and seawater spray.
Human particle-forming operations include combustion of fossil fuel in vehicles and power plants; Industrial activities such as mining, and combustion processes in various industries; Burning wood for heating and cooking, and burning vegetation (for example in felled agriculture and burning); As well as dust caused by soil and desertification due to non-sustainable farming.
On a global average, human sources of particulate matter are only about $10\%$ of the total particle in the atmosphere.
PM is divided into coarse particles, PM$_{10}$, and fine particles, PM2.5. The origin of the coarse PM$_{10}$ particles may be dust (from a natural or human source such as construction), agriculture, mining, fly ash, spores and plant particles. The source of the finer respiratory particles PM$_{2.5}$ is usually gases emitted as a result of fire and industrial combustion processes that become particles in the atmosphere.
During the COVID-19 period, PM$_{10}$ had increased relative to the closest period, over all the considered stations in Haifa and Gush-Dan. This, can be due to increasing in industrial pollution, as was noted above.
The increasing is on average 55.28 on Haifa, and 40.1 in Gush-Dan. The explanation percent of PM$_{10}$ by the period indicator is very small and between 0.01 to 0.03 over Haifa and Gush-Dan.
From the other hand, PM$_{2.5}$ had increased in most stations except for two stations, which had reduction. Over the stations with increasing in PM$_{2.5}$, the increasing in Haifa was 22.63 on average, and in 22.79 on average in Gush-Dan. For the two stations with reduction: the reduction in Haifa was -4.24 (general station), and in Gush-Dan it was -16.67 (traffic station). This is coincide with the fact that Gush-Dan is a traffic city, therefore the reduction is due to the reduction in transport, whereas the increasing in PM for most stations considered is due to increasing in industrial pollution.
The $R^2$ is at most $3\%$ over Haifa and Gush-Dan.
\subsubsection{SO$_{2}$}
Human sources of atmospheric sulfur dioxide emissions are the use of fossil fuels - mainly coal and oil, as well as in the smelting and production processes of metals and minerals from sulfur-containing lead. The most important human source of sulfur dioxide is the burning of sulfur-containing quartz fuels for home heating, electricity generation at power plants, and motorized vehicles.
In most stations in Haifa and Gush-Dan, SO$_{2}$ decreased during the COVID-19 relative to the closet period, but some stations had increasing in SO$_{2}$. The reduction is on average -28.31 on Haifa, and -5.22 in Gush-Dan. Over the stations with increasing in SO$_{2}$, the increasing is on average 17.14 in Haifa, and 5.99 in Gush-Dan. The explanation percent of SO$_{2}$ by the period indicator is very small and is 0.03 at most over Haifa and Gush-Dan.
\subsubsection{Volatile Organic Components}
Volatile organic compounds (VOCs) are organic chemicals with high vapor pressure at room temperature.
Man-made emission sources are divided into two main groups: combustion processes (air pollution from transport, industrial fuel combustion, vegetation fires and more), and evaporation processes (such as storage tanks, various products in the home or office and more). The main sources of man-made emissions can be divided into industrial, transport (and gas stations) and emissions in the home environment.
In our data, the pollutants Benzene, Toluene, and EthylB are components of VOCs, which were measured in one station in Haifa (general urban station). during the COVID-19, relative to the closet period, they all had reduction of -49.5, -25.65, and -55.83, respectively. The explanation percent of these three pollutants is 0.13, 0.01, and 0.04, respectively.

\subsection{Comparison of the full period over 2019-2020}
Here we compare the full period (January 1 until May 2) over 2019 and 2020. This comparison is in terms of the relative difference in pollutant amount only, without examining the period contribution to this difference. Each comparison is based on means comparison of a given pollutant at a specific station. The calculation of the mean omitted the missing values. The relative difference for a given pollutant at a specific station is defined as $(mean_{2020}-mean_{2019})/mean_{2019}$. Tables A1, 2 in the Appendix present the results. In addition, we used the t-test for two means comparison for testing the significance of the reduction/increasing for each pollutant. The resulted $p$-value is described in Tables A1, A2 in the Appendix. For testing the significant of the difference, we use the t-test for two means comparison.
A number of common trends are seen, as follows.
In Gush-Dan, significant reduction in PM$_{10}$ and PM$_{2.5}$ observed in all considered stations; significant reduction observed in the pollutants CO, NO$_{2}$, NO$_{x}$, and NO in all considered stations except for one station (Yad-Avner) which has no difference between 2019 and 2020.
In Haifa, significant reduction observed in the pollutants PM$_{10}$, PM$_{2.5}$, NO$_{2}$, NO$_{x}$,NO, and SO$_{2}$,  but a significant increasing observed in the pollutants O$_{3}$, EthylB, Toluene, and Benzene. These are measures for one station only.
The percent of change in Haifa is: NO$_{x}$: -28.09, NO: -36.24, NO$_{2}$:-24.95, SO$_{2}$: -23.09, PM$_{10}$: -25.96,  PM$_{2.5}$: -30.32, and in
Gush-Dan:  NO$_{x}$: -29.27, NO: -40.16, NO$_{2}$: -24.78, CO: -8.07, PM$_{10}$: -31.56, PM$_{2.5}$: -28.38.

%

\subsection{Comparison of the full period over the years 2000-2020}
\begin{landscape}
\begin{figure}[h!]
\bc
\includegraphics[width=2.4in, height=2.4in]{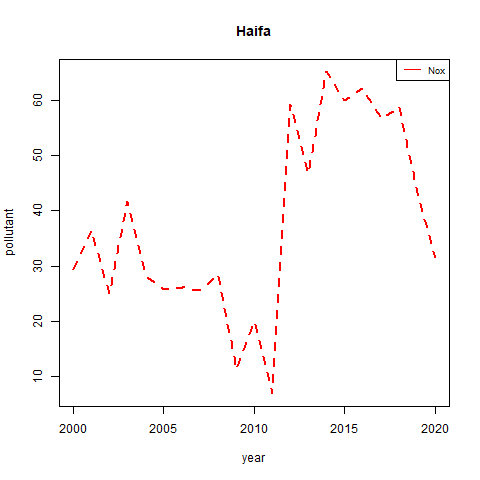}
\includegraphics[width=2.4in, height=2.4in]{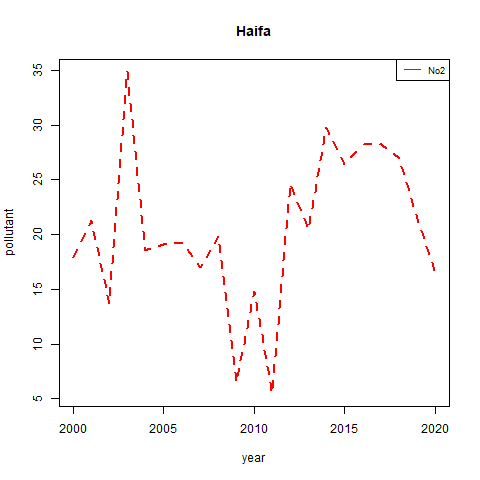}
\includegraphics[width=2.4in, height=2.4in]{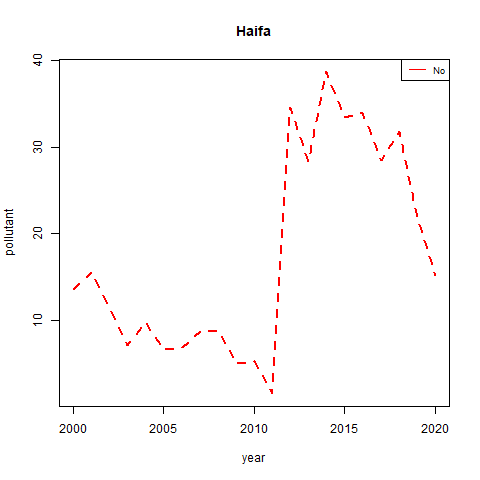}
\includegraphics[width=2.4in, height=2.4in]{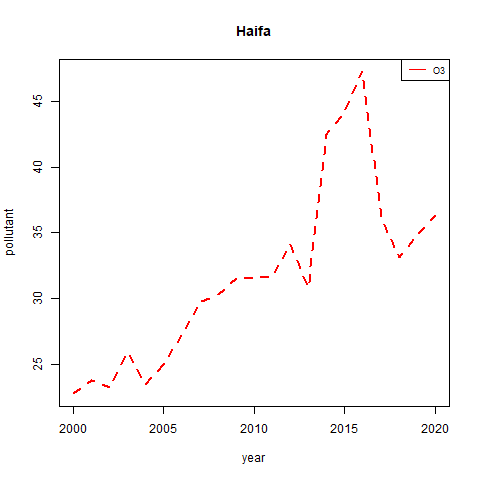}
\includegraphics[width=2.4in, height=2.4in]{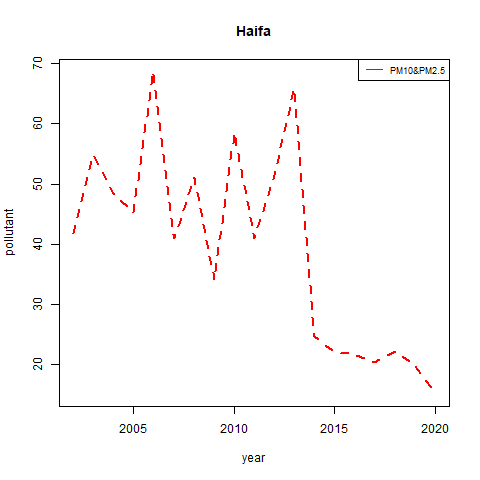}
\includegraphics[width=2.4in, height=2.4in]{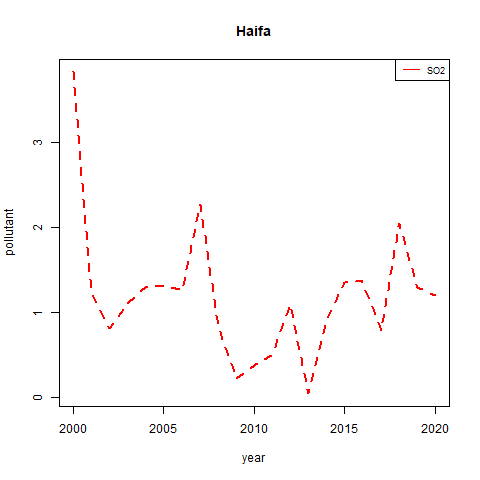}
\ec
\caption{Air pollution in Haifa over the years 2000-2020, in the period January 1 - May 2}
\label{fig:haifa_plots}
\end{figure}
\end{landscape}

\begin{landscape}
\begin{figure}[h!]
\bc
\includegraphics[width=2.4in, height=2.4in]{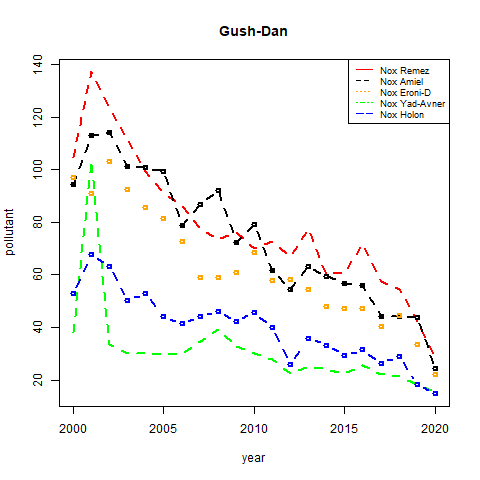}
\includegraphics[width=2.4in, height=2.4in]{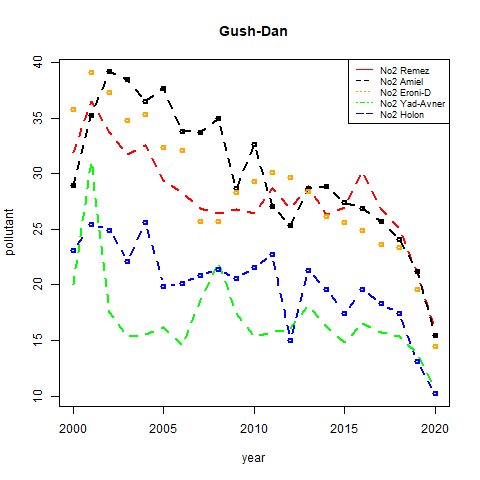}
\includegraphics[width=2.4in, height=2.4in]{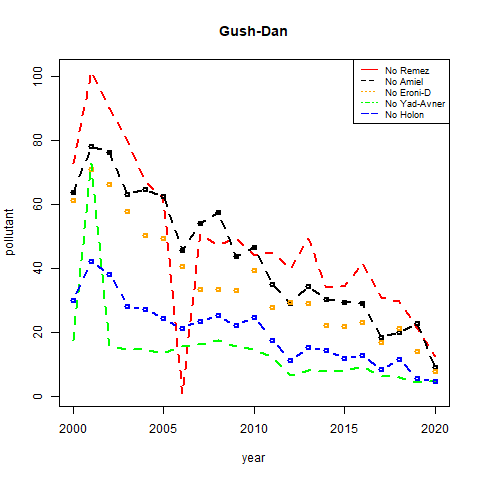}
\includegraphics[width=2.4in, height=2.4in]{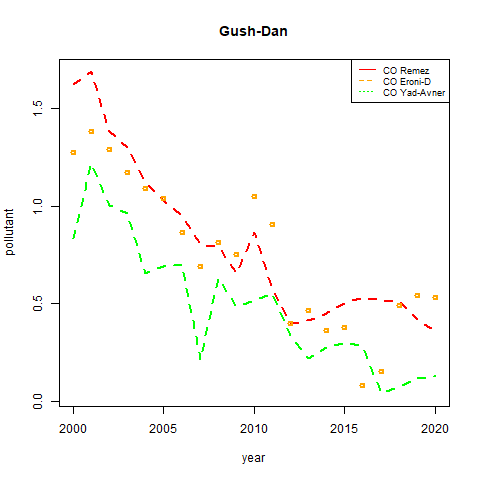}
\includegraphics[width=2.4in, height=2.4in]{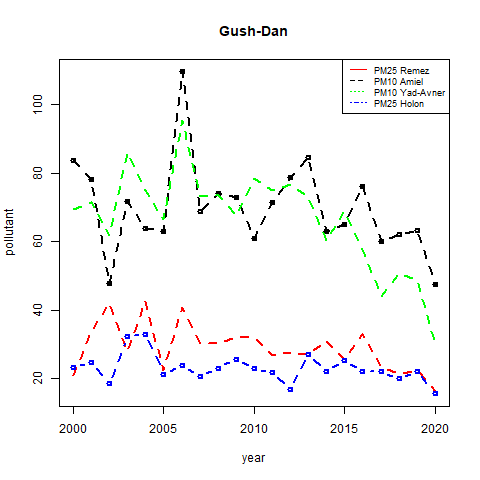}
\ec
\caption{Air pollution in Gush-Dan over the years 2000-2020, in the period January 1 - May 2}
\label{fig:GD_plots}
\end{figure}
\end{landscape}

In order to understand if the observed change between 2019 and 2020 during the full period (January 1 until May 2) is due to COVID-19 only, or it is  a part of a trend in the last years, we examined the behaviour of the pollutants in the full period over the years 2000-2020.
Figure\ \ref{fig:haifa_plots} and  Figure \ \ref{fig:GD_plots} present the trend lines for each pollutant at the various stations in Haifa and Gush-Dan, respectively. These trend lines are based on average of the measures at a specific year, where the calculation of the average omitted the missing values. Places in the graph where the trend line is truncated are due to missing values in the whole period over the specific year.
For Gush-Dan, the graphs are based on measures in the stations Remez, Amiel, IroniD, Yad-Avner, and Holon.
For Haifa, there was no unique station that had measures over the whole period over the years 2000-2020. Therefore the graphs are based on mixed of stations: for the years 2000-2010, the measures were taken from the station Market; for the years 2011-2020 the measures were taken from the station Azmaut, but from the station Park Carmel when no measures in station Azmaut, and from Nesher when no measures in the former stations. For the PM measures, there were some years with measures of PM$_{10}$, and some other years with measures of PM$_{2.5}$, therefore we combined these data into one graph. That is, the graphs presented for Haifa are only sketch to see the trend of each pollutant over the years 2000-2020.
\\

\indent The general trends that are seen in the graphs are as follows. For Gush-Dan, a decreasing trend is observed in NO$_{x}$, NO$_{2}$, No; the trend in PM$_{10}$ is zigzag, that is, jumping up and down; a decreasing trend in CO, but there is some increasing trend in the three recent years as observed in the stations IroniD and Yad-Avner.
For Haifa, an increasing trend is observed in NO$_{x}$, NO$_{2}$, and No started at the year 2011, but it had decreased in the last years; increasing trend in O$_{3}$, which had started to decrease at the year 2016, but still it is high; for PM, a decreasing trend started around the year 2013 and continue until now; some decreasing trend in SO$_{2}$ started around the year 2007, but there is some increasing trend in the last years. For all these pollutants in Haifa there is reduction in 2020, except of O$_{3}$ which increases.
Using Multiple comparisons of means (parametric t-test or non-parametric Wilcoxon test) for a given pollutant for each two consecutive years, some comparisons are significant and some of them are not. That is, during the years 2000-2020, we can see some years with a reduction in the amount of some pollutants.
Looking inside the activates that are related to transport and industry can explain these trends. The transport activities include a transfer of the diesel power plant to natural gas, and technological improvements in the engines and fuels of trucks, buses and cars. This resulted in  a marked decrease in sulfur oxides (SOx), and some decline in nitrogen oxides (NO$_{x}$). This also can explain the reduction of air pollution in Gush Dan.
However, there is no necessarily a reduction in PM and O$_{3}$.
In addition, improvement made in the fuels and in the combustion system in automotive engines, and catalytic converters had introduced. Therefore, the concentration of CO in the air had been reduced. Nevertheless, most exposure to high concentrations of CO occurs in areas where there is a dense concentration of slow moving vehicles, especially in busy city centers and along access routes.
\\
The reduction in air pollution from factories and power plants, especially in industrial pollution centres as Haifa Bay, is due to a combination of legislation, public pressure and increased awareness that lead to the adoption of cleaner technologies, cleaning and filtering technologies, beyond the use of natural gas.
As a summary, overall, there is a trend of reduction in air pollution in recent years. But although, the reduction that was observed in 2020 in some of the pollutants is even more than the decreasing trend seen before. That is, this expresses the influence of COVID-19 period, and its component of reduction transport pollution, and increasing of industrial pollution. All this true when assuming that there was no a significant change in weather, but it is mire or less the same over the recent years. That is, the change during the COVID-19 period is not part of the trend but the influence of the COVID-19 period.
\section{Impact of COVID-19 along with weather factors}
The contribution of the period indicator for the variation of the pollutant amount, as we saw in Section 3, is $22\%$ at most. But, needs to take into account other factors that may explain this variation. These factors can explain the difference in the pollutant amount during the COVID-19 period relative to its earlier closest period, as well. Due to the available data we have in hand, we examined the influence of the weather variables along with the indicator for period.
We did it by using a linear regression include the indicator variable \emph{ind} as in Section 3, and the weather variables wind-direction (WD), wind-speed (WS), temperature (Tmp), and relative-humidity (RH).
Because of possible correlations between these variables, the regression model includes usually some of these variables and not all of them together. That is, the model is of the form $Y=\alpha+\beta_1 ind+\sum _{j=2}^{p}\beta _{j} X_{j}$, where $Y$ is the amount of a given pollutant at a specific station, and $X_{j}$ denotes the weather variables.
The weather variables included in each model in addition to the period indicator are presented in Tables 6, 7 for Haifa and Gush-Dan, respectively. The  combination of the variables WD and WS is denoted in the tables as \emph{WDS}. These tables also report the $R^2$ for each model.
By this we get the percent of the total variation of the pollutant as explained by the variables included in the model. The maximum explanation percent of the total variation for each pollutant by the weather variables and the indicator for period together is as follows:
Nitrogen oxide (No, NO$_{x}$, and NO$_{2}$): 0.38 in Haifa, and 0.43 in Gush-Dan;
Ground-level ozone (O$_{3}$): 0.58 in Haifa, and 0.49 in Gush-Dan;
CO: 0.26 in Haifa, and 0.21 in Gush-Dan;
PM: 0.11 in Haifa, and 0.10 in Gush-Dan;
SO$_{2}$: 0.27 in Haifa, and 0.05 in Gush-Dan;
VOCs: 0.33.

\begin{landscape}
\begin{table}[h]
\caption{\textbf{Influence of COVID-19 and weather variables, period of January 1, 2020 - May 2, 2020, Haifa}}.
\centering
\fontsize{7.4}{1.16}\selectfont
\begin{tabular}{c|c|c|c|c|c|c|c|c|c|c|c|c|c|c|c}
\midrule
&Regavim&Shprintzak&Hogim&Ahuza&Igud&Hasidim&Shaanan&Nesher&Carmel&Ata&Binyamin&Tivon&Yam\\\\
\midrule
NO$_{2}$&wds&wds&wds&wds,tmp&wds&wds&wd,rh&tmp&wds,tmp&ws,rh,tmp&wds&wds,tmp&wds,rh\\
\midrule
$R^2$&0.36&0.22&0.30&0.20&0.20&0.04&0.06&0.03&0.26&0.18&0.10&0.14 &0.38\\
\midrule
\midrule
No&wds&wds&wds&wds&wds&wds&wds,rh&rh&wds,tmp&ws,rh&wds&wds,tmp&wds,rh,tmp\\
\midrule
$R^2$&0.07&0.03&0.10&0.08&0.04&0.00&0.03&0.02&0.04&0.03&0.01&0.09&0.13 \\
\midrule
\midrule
NO$_{x}$& wds&wds&wds&wds,tmp&wds&wds&wds&rh&wds,tmp&ws,rh&wds&wds,tmp&wds,rh,tmp\\
\midrule
$R^2$&0.28&0.19&0.29&0.18&0.17&0.03&0.15&0.03&0.23&0.14&0.09&0.14&0.31\\
\midrule
\midrule
SO$_{2}$&wds&&wds&wds&wds&wds&wds,rh&rh,tmp&wds,rh&ws,rh&wd&wds,tmp\\
\midrule
$R^2$&0.06&&0.11&0.06&0.04&0.01&0.16&0.14&0.27&0.13&0.02&0.15\\
\midrule
\midrule
O$_{3}$&&wds&wds&&wds&wds&wds,rh,tmp&rh,tmp&wds,rh,tmp&wds,rh&&&wds,rh\\
\midrule
$R^2$&&0.20&0.29&&0.40&0.33&0.33&0.22&0.35&0.53&&&0.58\\
\midrule
\midrule
CO&&&&&&&ws,rh&&wds,rh\\
\midrule
$R^2$&&&&&&&0.13&&0.26\\
\midrule
\midrule
PM$_{2.5}$&wds&&&wds&wds&&wds,rh,tmp&rh,tmp&wd,rh&wds&wds&wd,tmp&wds,rh\\
\midrule
$R^2$&0.06&&&0.05&0.05&&0.11&0.07&0.06&0.03&0.04&0.05&0.10\\
\midrule
\midrule
PM$_{10}$&wds&&&wds,tmp&wds&&wds,rh&rh&&wds,rh&wds&wds\\
\midrule
$R^2$&0.02&&&0.06&0.04&&0.03&0.02&&0.03&0.02&0.03\\
\midrule
\midrule
Benzene&wds&&&&wds&&&&&&wds\\
\midrule
$R^2$&0.33&&&&0.31&&&&&&0.08\\
\midrule
\midrule
Toluene&wds\\
\midrule
$R^2$&0.15\\
\midrule
\midrule
EthylB&wds\\
\midrule
$R^2$&0.19&\\
\midrule
\bottomrule
\end{tabular}
\end{table}
\end{landscape}

\begin{landscape}
\begin{table}[h]
\caption{\textbf{Influence of COVID-19 and weather variables, period of January 1, 2020 - May 2, 2020, Gush-Dan}}.
\centering
\fontsize{7.4}{1.16}\selectfont
\begin{tabular}{c|c|c|c|c|c|c|c|c|c|c|c|c|c|c|c}
\midrule
&Komemiyut&Yoseftal&Avner&Holon&PT\\
\midrule
NO$_{2}$&rh,tmp&rh,tmp&rh,rain&wd,rh&wds\\
\midrule
$R^2$&0.12&0.13&0.13&0.11&0.43\\
\midrule
\midrule
No&rh,tmp&rh,tmp&rh,rain&wd,rh&wds\\
\midrule
$R^2$&0.10&0.12&0.03&0.02&0.15 \\
\midrule
\midrule
NO$_{x}$& rh,tmp&rh,tmp&rh,rain&wd,rh&wds\\
\midrule
$R^2$&0.10&0.12&0.09&0.07&0.30\\
\midrule
\midrule
SO$_{2}$&&&rh&&wds\\
\midrule
$R^2$&&&0.05&&0.02\\
\midrule
\midrule
O$_{3}$&&&rh,rain&&wd\\
\midrule
$R^2$&&&0.23&&0.49\\
\midrule
\midrule
CO&&&rh,tmp\\
\midrule
$R^2$&&&0.21\\
\midrule
\midrule
PM$_{2.5}$&tmp\\
\midrule
$R^2$&0.10\\
\midrule
\midrule
PM$_{10}$&&&rh,tmp\\
\midrule
$R^2$&&&0.05\\
\midrule
\bottomrule
\end{tabular}
\end{table}
\end{landscape}

\section{Discussion}
The period of COVID-19 breakout had influenced the air pollution in Israel in two components, pollution from transport and pollution from industrial. Relative to its earlier closest period, the COVID-19 period caused a reduction in pollution from transport, and an increasing in pollution from industry due to increasing in infrastructures building and home renovations. We evaluate the pollution from transport  mainly by the Nitrogen oxide (i.e., NO$_{x}$, NO$_{2}$, and No), and the pollution from industrial by O$_{3}$. Over the regions of Haifa and Gush-Dan, the reduction in pollution from transport was on average between -33$\%$ to -46$\%$, while the increasing in O$_{3}$ was on average, between 21$\%$ to 46$\%$. Other pollutants had the both changes of increasing and decreasing over the various considered stations. That is, they captured the both influences of transport and industry. But overall, all pollutants except the VOCs decreased relative to the same period in the last year. The influence of the period over the total variation of each pollutant is at most 22$\%$, while adding weather variables resulted in explanation percent of 58$\%$ at most.
As a conclusion, although the sharp reduction in transport, still the reduction in air pollution from it was less than 50$\%$, and the percent of explanation by period was not high. Adding the weather variables increases the percent of explanation of the total variation of each pollutant, for a level which is higher than 50$\%$. That is, there is still high percent that we was not explained.
Comparing with the previous period, reduction was in almost all pollutants, that is, the increasing in industrial not achieve the general reduction percent in pollution form industrial. But still the reduction was less than 50$\%$.
Two possible explanations for this. First, there are other possible factors that may influence the air pollution, which we did not control them. Second, even if we completely stop polluting, probably some of the pollution will sink but some will remain. That is, the decay in pollution may be non-linear, and therefore should try a polynomial relation for further research.

\vspace{6pt}

%
%


\newpage


\newpage
\begin{landscape}
\section*{Appendix}
The following tables presents air pollution during COVID-19 period (period of January 1 , 2020 - May 2, 2020) in 2020 comparing to 2019.

\begin{table}[h!]
\caption{\textbf{Haifa Stations}}.
\centering
\fontsize{6.84}{1.16}\selectfont
\begin{tabular}{c|c|c|c|c|c|c|c|c|c|c|c|c|c}
\midrule

Station&Azmaut     &Regavim&Regavim &Park Carmel&Regavim&Azmaut&Regavim&Azmaut&Regavim&Azmaut&Regavim&Nesher
\\
\midrule
Pollutant&PM$_{10}$& PM$_{10}$& PM$_{2.5}$& SO$_{2}$&  SO$_{2}$ &NO$_{x}$ &NO$_{x}$ &NO&NO&NO$_{2}$ &NO$_{2}$&O$_{3}$\\
\\
\midrule
2019&19.17&		49.10&19.79&1.29&0.74&	43.32&	     12.16&	    21.89&2.25&	21.48&9.94&	34.90\\
\midrule
2020&14.94&	34.80&13.79&1.20&0.45&	31.60&	           8.53&		15.17&1.31&	16.53&7.27&	36.35\\
\midrule
Relative Diff&-22.07&	-29.85&-30.32&-6.98&-39.19 &-27.05&	-29.12&	-30.70&-41.78&-23.04&-26.86&4.15 \\
\midrule
$p$-value&$<$ 2.2e-16&$<$ 2.2e-16&$<$ 2.2e-16&2e-08&   $<$ 2.2e-16&  $<$ 2.2e-16&	$<$ 2.2e-16&$<$ 2.2e-16&$<$ 2.2e-16&	$<$ 2.2e-16&	$<$ 2.2e-16&$<$ 2.2e-16 \\
\bottomrule
%
\midrule

Station&Regavim&Regavim&Regavim\\
\\
\midrule
Pollutant&EthylB&Toluene&Benzene\\
\\
\midrule
2019&	0.14&0.41&	0.13\\
\midrule
2020&	0.23&0.61&	0.15\\
\midrule
Relative Diff&	64.29&48.78&	15.38\\
\midrule
$p$-value&$<$ 2.2e-16&$<$ 2.2e-16&$<$ 2.2e-16 \\
\bottomrule
\\
\end{tabular}
\end{table}
\end{landscape}

 \begin{landscape}
\begin{table}[h]
\caption{\textbf{Gush-Dan Stations}}.
\centering
\fontsize{6.84}{1.16}\selectfont
\begin{tabular}{c|c|c|c|c|c|c|c|c|c|c}
\midrule

Station&Amiel&Yad-Avner&Holon&Remez&Remez&IroniD&Avner\\
\midrule
Pollutant&PM$_{10}$&PM$_{10}$&PM$_{2.5}$&PM$_{2.5}$&CO&CO&CO\\
\midrule
2019&	63.23&48.63&22.23&	22.54&	0.42&	0.54&	0.12 \\
\midrule
2020&	47.44&	30.08&	15.71&	16.36&	0.36&	0.53&	0.13\\
\midrule
Relative Diff &-24.97&	-38.15&	-29.33&	-27.42&	-14.29&	-1.85&	8.33 \\
\midrule
$p$-value&$<$ 2.2e-16&$<$2.2e-16&$<$2.2e-16&$<$2.2e-16&$<$2.2e-16&0.055&	2.6e-07 \\
\bottomrule
%
\midrule

Station&Remez &Amiel&IroniD&Avner &Holon&	Remez&Amiel\\
\midrule
Pollutant&NO$_{2}$& NO$_{2}$&NO$_{2}$&NO$_{2}$&NO$_{2}$&NO$_{x}$&NO$_{x}$\\
\midrule
2019&21.00&	21.23&	19.57&	13.83&	13.11&	42.13&	43.94 \\
\midrule
2020&15.76&	15.42&	14.43&	10.62&	10.21&	28.31&	24.55	\\
\midrule
Relative Diff&-24.95&	-27.37&	-26.26&	-23.21&	-22.12&	-32.80&	-44.13  \\
\midrule
$p$-value&$<$ 2.2e-16&	$<$ 2.2e-16&	$<$ 2.2e-16&	$<$ 2.2e-16&	$<$ 2.2e-16&	$<$ 2.2e-16&	$<$ 2.2e-16 \\
\bottomrule
%
\midrule

Station&IroniD &Yad-Avner &Holon&Remez&Amiel&IroniD &Holon&Yad-Avner\\
\midrule
Pollutant&NO$_{x}$&NO$_{x}$&NO$_{x}$&NO&NO&NO&NO&NO\\
\midrule
2019&33.61&18.27&18.37&21.16&22.72&14.03&5.58&4.44 \\
\midrule
2020&22.20&	15.27&	14.87&	12.67&	9.12&	7.81&	4.67&	4.66\\
\midrule
Relative Diff&-33.95&-16.42&-19.05&-40.12&-59.86&-44.33&-16.31&4.95 \\
\midrule
$p$-value&$<$ 2.2e-16&	$<$ 2.2e-16&	$<$ 2.2e-16&	$<$ 2.2e-16&	$<$ 2.2e-16&$<$ 2.2e-16&0.0005&0.1055 \\
\bottomrule
\\
\end{tabular}
\end{table}
 \end{landscape}




\end{document}